\documentclass[sigconf]{acmart}

\AtBeginDocument{%
  }
\usepackage{float}
\usepackage{url} 
\providecommand{\myparabb}[1]{\smallskip\noindent\textbf{#1} }

\begin{document}

\title{Visual Studio Code in Introductory Computer Science Course: An Experience Report}

\author{Jialiang Tan}
\email{jtan02@wm.edu}
\affiliation{%
  \institution{William \& Mary}
  \state{Virginia}
  \country{USA}
}

\author{Yu Chen}
\email{ychen39@wm.edu}
\affiliation{%
  \institution{William \& Mary}
  \state{Virginia}
  \country{USA}
}

\author{Shuyin Jiao}
\email{sjiao2@ncsu.edu}
\affiliation{%
  \institution{North Carolina State University}
  \state{North Carolina}
  \country{USA}
}

\begin{abstract}
Involving integrated development environments (IDEs) in introductory-level (CS1) programming courses is critical. However, it is difficult for instructors to find a suitable IDE that is beginner friendly and supports strong functionality. In this paper, we report the experience of using Visual Studio Code (VS Code) in a CS1 programming course. We describe our motivation for choosing VS Code and how we introduce it to students. We create comprehensive guidance with hierarchical indexing to help students with diverse programming backgrounds. We perform an experimental evaluation of students' programming experience of using VS Code and validate the VS Code together with guidance as a promising solution for CS1 programming courses.


\end{abstract}

\maketitle

\section{Introduction}

Integrated Development Environments (IDEs) play an important role in learning a programming language. IDEs offer programmers extensive development abilities. They understand language syntax and provide features such as build automation, code linting, testing, and debugging, which accelerate and simplify the coding process. 
Through the help of IDEs, students benefit from efficient programming, testing, and debugging. Students can further develop better coding habits and flatten the learning curve of a new language.
As a result, more and more instructors start involving IDEs in introductory-level Python courses such as Atom~\cite{atom}, Jupyter Notebook~\cite{perez2015project, al2022jupyter, van2020jupyter}, and many others, which significantly improve student's coding experiences.

However, not all IDEs are suitable for introductory-level Python courses. Choosing a satisfying IDE is difficult for instructors. 
On the one hand, professional IDEs~\cite{vim, eclipse, intelliJ} support an integrated programming environment and many powerful features but with limited support for education. One may need advanced knowledge and plenty of time to use them properly. For students who are new to programming, it is hard to take advantage of IDEs on top of learning a new programming language, because plenty of time is spent on software installation and environment setup. On the other hand, education-focused code editors are easily installed and manipulated, but they are rarely used during professional software development cycles. Students face a big gap when transiting from college to industry~\cite{valstar2020faculty, valstar2020quantitative}, which has some negative impacts on their future careers.

\sloppy
Thus, there is urgent to use an appropriate IDE for the introductory-level Python class, which not only provides enough education-related features but also is mainstream among professional software engineers. We aim to use an IDE that satisfies four features: 1) it is cross-platform supported, students are able to install and use it identically on different OS/hardware; 2) it is easy to use, students can code effectively with a simple and concise interface; 3) it has extensive functionalities, e.g. code compilation, project organization, and multi-languages support, etc.; 4) it is mainstream among professional software engineers, which means the IDE is widely used in industry, bridging the gap between introductory-level class and future career.

Thus, we identify Microsoft Visual Studio Code (VS Code) as the desired IDE, which has all four aforementioned features. Moreover, VS Code has been adopted in many advanced courses in our department, such as operating systems, compiler constructions, computer networks, and many others.
However, VS Code does not draw significant attention to CS1 courses. Furthermore, with our study of 20 computer science departments, none of them specify VS Code as the default IDE in the introductory Python courses. Moreover, there is no comprehensive guidance on VS Code for educators and students. 
Since students have various backgrounds, e.g. from diverse majors, with uneven learning speeds, or are familiar with different operating systems, comprehensive guidance is necessary to facilitate the efforts of both students and educators. 

In this paper, we give the first experience report for the use of VS Code in an introductory (CS1) Python course. We create the first comprehensive guidance of VS Code for both students and educators by gathering plugins (i.e., VS Code extensions) for Python education and various user experiences. The guidance is highly modularized. Students who have programming experience jump to the sections they need to learn and skip the sections they already know; students who are beginner programmers follow the guidance step-by-step; students who come across errors target why errors occur and how to solve them. With the help of our guidance, students can quickly acquire the necessary knowledge of the programming environment in the class. We already integrate VS Code and our guidance into the introductory-level Python course in our department. We use a survey to evaluate the effectiveness of our efforts. From the survey, the students highly value the use of VS code and our guidance. Students with some industry internship experiences acknowledge the usefulness of learning Python with VS Code over other education-oriented IDEs, as VS Code is widely used in the industry.

\myparabb{Contributions.} We make the following four contributions. 

\begin{itemize}
    \item We point out the importance of selecting a proper IDE for introductory-level courses.  
    \item We identify Visual Studio Code as a desired IDE. We investigate it from four aspects and analyze its practicality for introductory-level Python courses.
    \item We propose the first comprehensive guidance\footnote{The VS Code guidance will become public upon this paper's acceptance.} with hierarchical indexing, to guide students with diverse backgrounds.
    \item We report our experiences in using VS Code in an introductory-level programming course and show that VS Code is a satisfactory IDE for the introductory-level Python course. We also verify the value and necessity of the guidance.

\end{itemize}

\myparabb{Organization.} Section 2 reviews the background and related work. Section 3 shows our investigation of VS Code. Section 4 describes the guidance and support we offer to students. Section 5 shows some experimental studies. Section 6 provides the discussion. Section 7 presents conclusions and future work.




\section{Background and Related Work}

\subsection{Python in Education}


Python has become the most prominent language in college, it tops the list of the most-taught programming languages, especially in introductory computer science courses, it is mainly due to three facts: {\it Python is beginner friendly}, it has a simplified syntax with an emphasis on natural language while most programming languages have complex rules, thus it is perfect for beginners to learn and practice~\cite{khoirom2020comparative, pellet2019beginner, srinath2017python, cutting2021review}. 
{\it Python is a powerful tool}, it is applied in many areas like machine learning and big data, especially effective and essential in scientific computing and data analysis~\cite{oliphant2006guide, van2011numpy, scikit-learn, chen2020atmem, chen2020efficient, tan2021toward, tensorflow2015-whitepaper, paszke2017automatic}. Students with diverse backgrounds use Python for different purposes, e.g. analyzing the market in finance, simulating protein structure in biology. Students who pursue a formal education in computer science or computer science-related majors are extremely likely to continue using Python throughout their careers.  
{\it Python is popular}, according to the PYPL index~\cite{pypl} and Stackoverflow Developer Survey 2021~\cite{survey}, Python is the most popular and the fastest-growing programming language (10.4\%) in the world as of Jun 2022.



\subsection{Pedagogical Approaches}
Programming language plays an important role in computing education. Many systematic studies explore the teaching in introductory-level programming classes~\cite{pears2007survey, schulte2006teachers, medeiros2018systematic, salleh2010empirical}. 

Through our observation, a tremendous number of research have been done on how the programming environment can improve students' coding experience in recent years, especially the programming environment for Java and Python, which are the most taught entry-level programming languages in college. 
The selection of the programming environment is typically decided by instructors. Pedagogical tools and Intelligent Tutoring Systems (ITS) ~\cite{pythontutor, van2004teaching, vanlehn2011relative, crow2018intelligent, naser2008developing} have been designed specifically for programming novices. Also, some works report the study of using industry-level programming environments such as Jupyter Notebooks~\cite{al2022jupyter, depratti2019using, barba2019teaching, van2020jupyter} or Eclipse extension~\cite{reis2004taming}, their experience shows how a great IDE improve students' engagement and performance.

\subsection{IDEs in Education}
Integrated Development Environments (IDEs) refer to software applications that combine all tools needed for a software development project, including an editor, compiler, and debugger. They consolidate different aspects of software development and significantly improve programmers' productivity.

Many types of programming environments exist in the market. Roughly there are two kinds: plain-text/code editor and IDE. Text/code editor does not require complex installation and configuration, some editors are installed by default like NotePad or TextEdit, but they offer limited functionalities and are not directly related to programming. On the other hand, a full-featured IDE combines functionalities in one, integrating all tools developers need to build and test, but IDE requires more disk space/memory or a faster processor, users may suffer more from installation, configuration, even cost (some IDEs' license are expensive). However, there is not a clear boundary exist between them. Users may turn a text/code editor into an IDE by installing plug-ins/extensions. There are always trade-offs between the time spent on installation/configuration and how 
powerful the functionalities are.



It is important to introduce IDEs in programming courses, especially in introductory-level courses. 
A great IDE helps students from two perspectives. 
First, it helps students write correct code. In our past teaching experience, common errors come up over and over again in students who have no or less programming experience. 
Second, it helps students establish good coding habits. Beginners focus more on correctness and tend to produce low-quality code, including messy code format, meaningless variable names, excessive function length, etc. Many IDEs support features like code formatting, variable name suggesting and function length warning,  helping students write clean and decent code that shows professionalism towards industry standards.




\section{Visual Studio Code Investigation}
\label{investigation}

Visual Studio (VS) Code is a lightweight IDE developed and supported by Microsoft, it is free for private or commercial use. The core feature of VS Code is its extension support, users are able to add languages, debuggers, and tools to their own installation to better serve later development. Besides standard extensions released by Microsoft, there are plenty of extensions on the VS Code Extension Marketplace~\cite{marketplace} contributed by third-party organizations and individual developers. We analyze its practicality for introductory-level Python courses from four aspects: accessibility, easy-to-use, functionality, and popularity.

\myparabb{Accessibility. } VS Code provides cross-platform support, it runs on various operating systems, i.e. macOS, Windows, Linux, and on most available hardware with an identical user interface. VS Code has a small download (< 200 MB) and a disk footprint (< 500 MB), it is lightweight, and perfectly fits every student with different devices. 

As a multi-languages supported IDE, almost every major programming language has extension support, it is effortless to switch between different programming languages. Though our class is for Python beginners, students can continually code with VS Code in later programming classes.

\begin{figure}[t]
    \centering
    \includegraphics[width=0.5\textwidth]{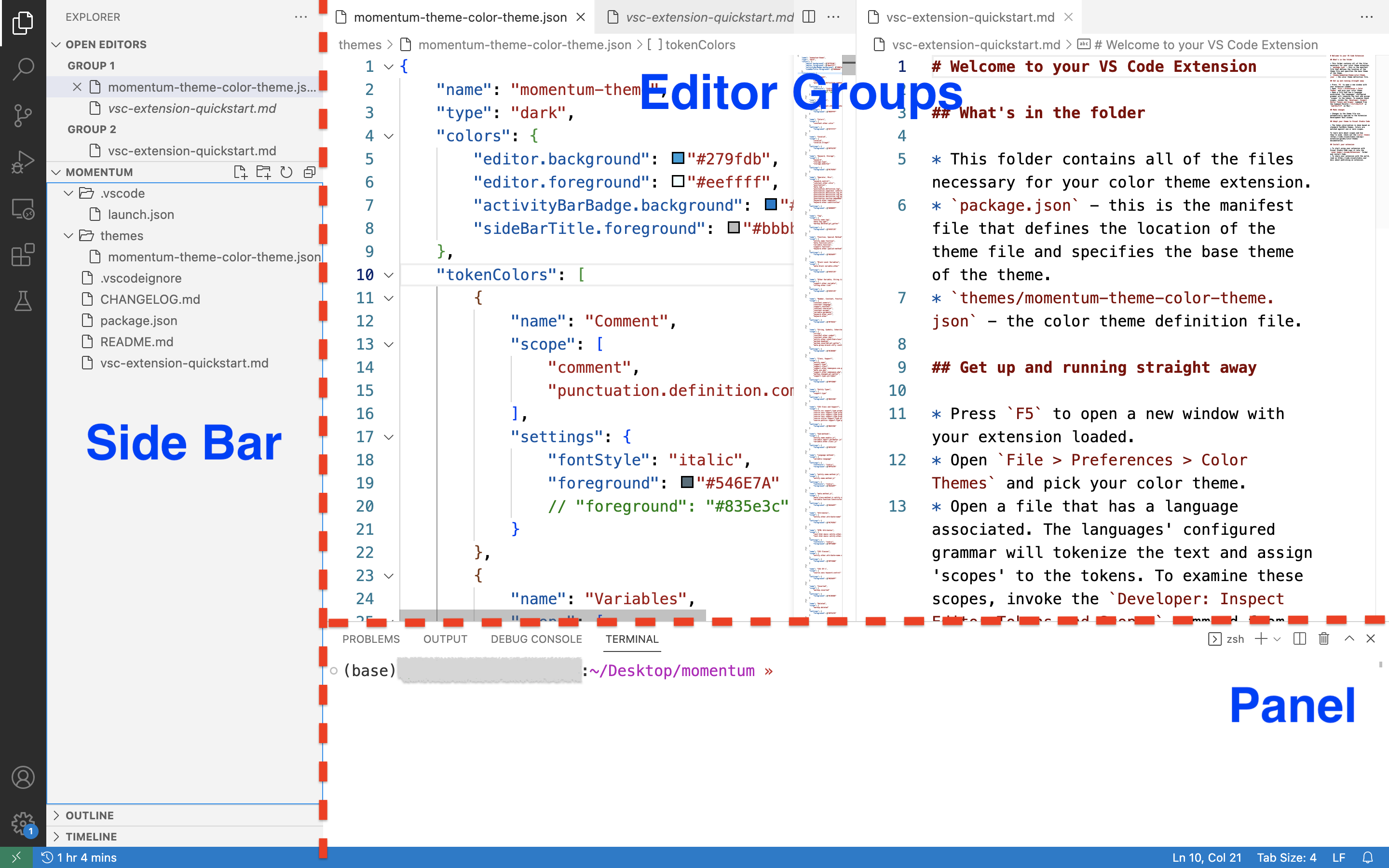}
    \caption {The basic layout of VS Code user interface~\cite{userinterface} that we introduced to students. The editor is the area to edit files, students can open multiple editors at the same time side by side vertically or horizontally. The panel below the editor is for output or debug information, errors and warnings, or an integrated terminal. The side bar contains different views like the Explorer or Extension Marketplace, to assist students while working on the projects or downloading extensions.}
    \label{fig:interface}
\end{figure}

\myparabb{Easy-to-use. } VS Code has a compact but simple user interface. Figure~\ref{fig:interface} shows an example. In the center is the editor, where students code their assignments. Under the editor is the panel, it can be displayed in different panels like a debug console or a built-in terminal, the terminal always starts from the root of a certain workspace. On the left side is the sidebar that contains different views, Figure~\ref{fig:interface} shows the view of Explorer to better assist in locating a file. Maintaining a single window of code and debugging effectively increase programming efficiency.

Besides the concise user interface, VS Code employs many features to improve development. One important feature that simplifies programming is the workspace configure setting. A VS Code workspace is the root folder of the current project, configure settings that apply to a specific workspace but not others. As a common scenario, students need to have different settings (e.g. interpreter version, dependent libraries, programming languages) among projects. In our past experience, students may feel confused with the configure settings, especially set/reset environment at the beginning of starting a new project. It will be much easier to work on VS Code workspace, because it allows configuring settings in the context of the current workspace, and always overrides the global user settings. 

\begin{figure}[t]
    \centering
    \includegraphics[width=0.5\textwidth]{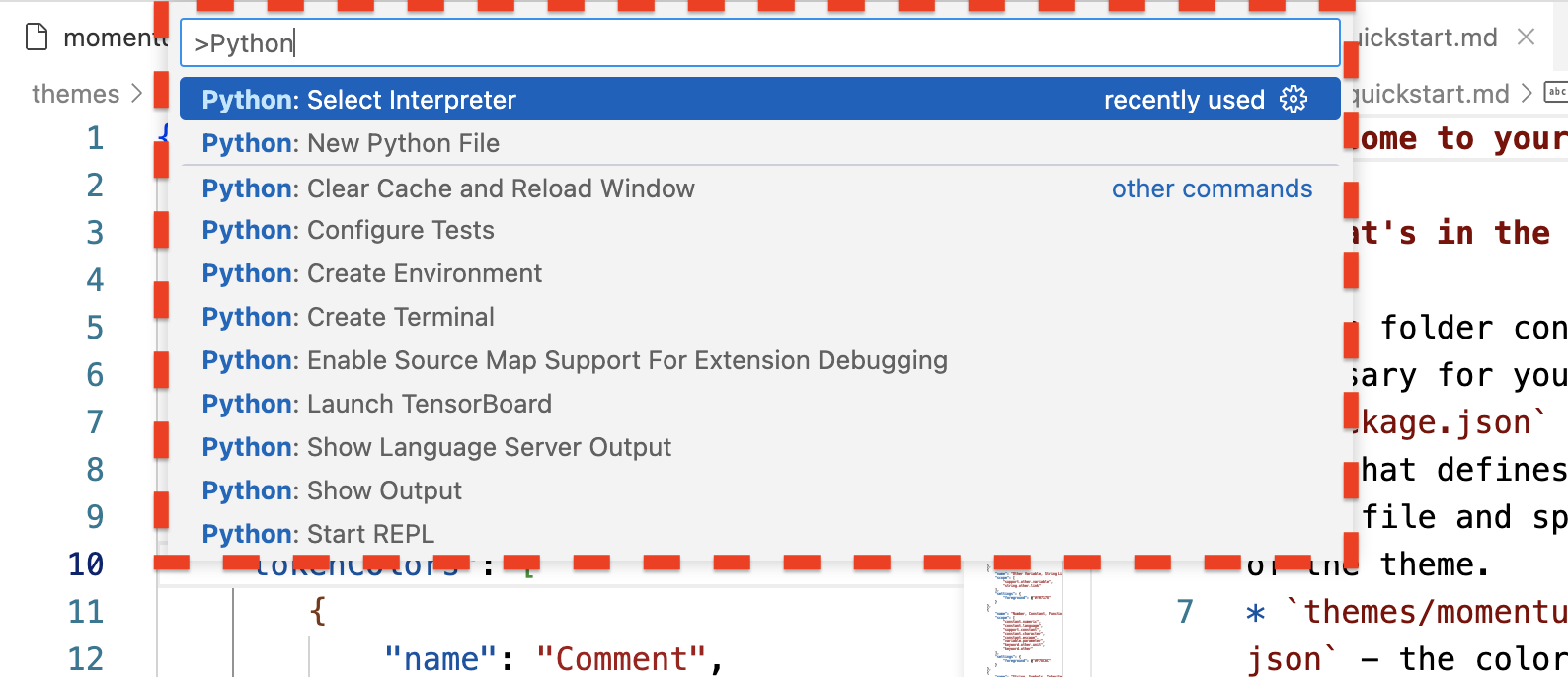}
    \caption{The red box indicated the Command Palette of VS Code. Students can access all functionality of VS Code and installed extensions through the Command Palette.}
    \label{fig:command}
\end{figure}

Furthermore, VS Code supports remote development. The {\it Remote SSH} extension allows opening a remote folder on any remote machine, virtual machine, or container a running SSH server. Another useful feature favored by many developers is the {\it Command Palette}, Figure~\ref{fig:command} shows an example, where users would have access to all of the functionalities of VS Code, including commands of your installed extensions.  

\begin{figure}[t]
    \centering
    \includegraphics[width=0.5\textwidth]{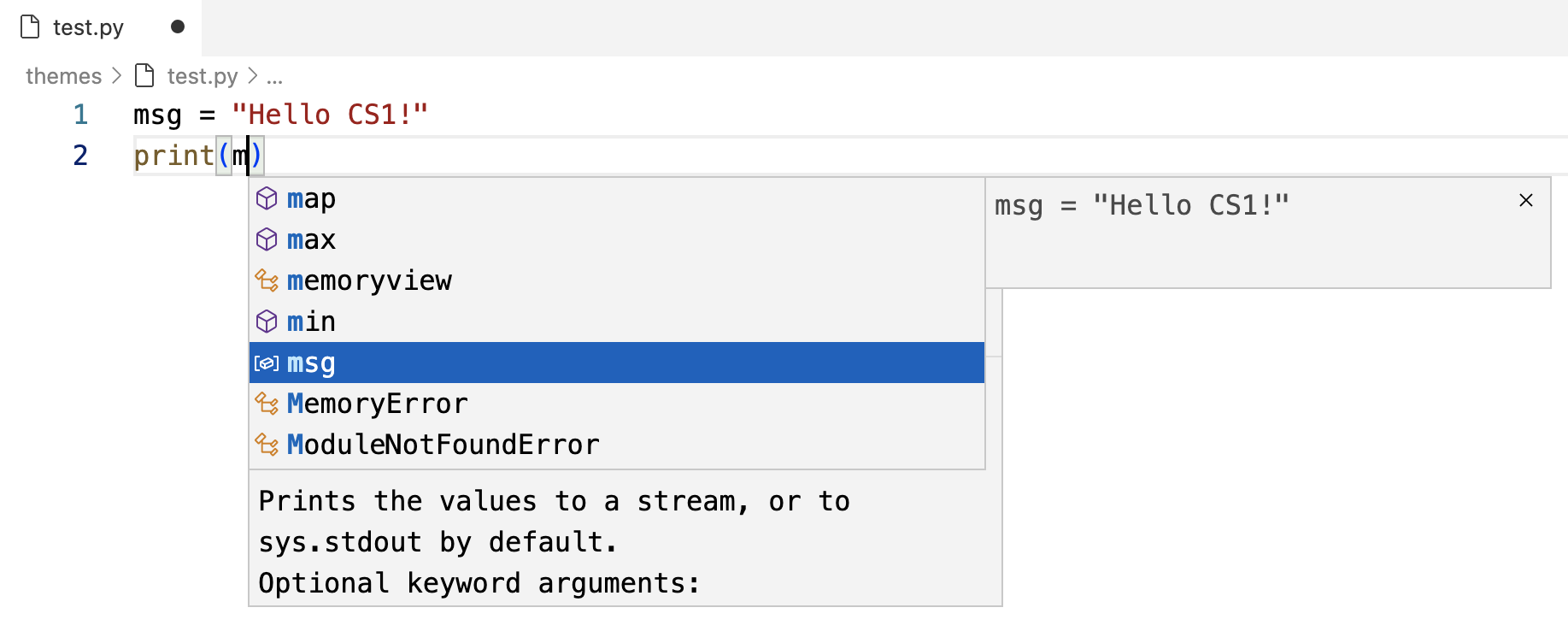}
    \caption{An example of code completion supported by VS Code, where the IntelliSense suggests using variable {\tt msg} in {\tt print} function.}
    \label{fig:intellisense}
\end{figure}

\myparabb{Functionality. } Besides the basic features of a source code editor, VS Code offers extensions to increase its functionality. With over 30,000 extensions in Marketplace, users pick favored extensions and customize their installation to improve the programming experience. For example, in our Python course, we recommend students install the Python extension developed by Microsoft, which offers strong support for Python language, it has powerful features such as IntelliSense~\cite{intellisense}, code formatting, debugging, variable explorer, and more. The IntelliSense suggestion pops out while you type, it provides intelligent code completion based on the language semantics and written source code. Figure~\ref{fig:intellisense} shows an example, where IntelliSense suggests using variable {\tt msg} in {\tt print} function.

\myparabb{Popularity. } VS Code is widely used among real-world developers. In the Stack Overflow 2021 Developer survey~\cite{survey}, VS Code tops the most popular developer environment tool, with 71.06\% of over 80,000 respondents reporting that they use it. Due to the great user amount, bugs/issues are fixed/solved in time. VS Code updates frequently, from June 2020 to May 2021, VS Code development team made 22 releases in total, updating almost every month. 

The high popularity tops the reasons why we introduce VS Code to our Python class. Students encounter difficulties in switching IDEs, they are able to use VS Code throughout different programming languages classes, or in their future careers. 

\section{Guidance and Support}

To accelerate the learning process of VS Code for students with various backgrounds, we introduce comprehensive guidance in Python class. The guidance\footnote{Guidance figures are not shown in the paper for anonymous purposes, and the guidance will become public upon this paper's acceptance.} is divided into separate tutorials, the following subsections explain the detail of each tutorial.

\myparabb{Download and installation of VS Code.}
In this tutorial, students are guided to the download page, and how to choose the correct version according to their OS. We introduce the workspace right after the installation because it is the fundamental unit to manipulate projects. We give examples of how to start a VS Code project, and how to organize files in a project.




\myparabb{Download and installation of Python.}
In our past teaching experience, we suffer three main problems with Python version: 1. not being aware of the difference between Python2 and Python3; 2. installing the wrong version; 3. having both Python2 and Python3 installed on the same device, but don't know how to switch them.


In the tutorial, we require students to use Python3 and emphasize the difference between Python2 and Python3. Before the installation, students need to check if they have any versions of Python pre-installed. If they don't have or have Python2 installed, they are led to the newest Python3 download link. If they have any version of Python3 installed, then there is no need to install it again. For students with multiple versions of Python installed, we provide a detailed guide on switching Python versions in the next tutorial.

\myparabb{Set up Python environment.}
Next, students install the Python extension and set up the Python environment in VS Code. In our past teaching experience, students were always confused with the programming environment settings, we pay extra effort into where and how to set/change configure settings. 

Then students are guided to the extension marketplace, to find and install the Python extension developed by Microsoft. The Python extension works on/with any operating system/Python version, the Jupyter extension is included in the Python extension installation bundle.
The Python extension supports code completion and IntelliSense on top of the currently selected Python interpreter, we provide step-by-step guidance with figures on how to select a Python interpreter from {\it Command Palette}. We also provide guidance on how to change programming language mode, for students who want to work with other programming languages in later classes.



\myparabb{Run Python examples. }
Once set up Python environment, VS Code becomes a real Python IDE. In this tutorial, we guide students to write, run their first simple Python program from scratch, and give explanations on how Python extension improves coding. We require students to generate expected outputs for example programs, such that they have identical settings for the Python environment. 

\myparabb{Install and run Jupyter extension.} We use VS Code's Jupyter extension to run in-class coding exercises to improve student engagement~\cite{perez2015project, al2022jupyter}, through the Jupyter extension students open and run ipynb files on VS Code the same way as Python files. The Jupyter Notebook extension provides basic notebook support for language kernels and allows any Python environment to be used as a Jupyter kernel. In this tutorial, we guide students to install the Jupyter extension, create new Jupyter files, and manage and run the code cells.

\myparabb{Install and use scientific packages.} Python Packages are very efficient to solve complex problems in scientific computing, data visualization, data manipulation, and many other fields. In this tutorial, we guide students to install and import Python libraries, e.g., {\tt Numpy}, {\tt Pandas}, {\tt Matplotlib}, and {\tt SciPy}. Then students are required to learn and use basic APIs, and the example codes are provided. To succeed in this tutorial, students need to plot a certain figure using {\tt Matplotlib} functions.

\subsection{Improvements for Guidance}
\label{guidanceimp}

To better guide students, the guidance comprises optimizations as follows:  

\begin{itemize}
    
    \item We provide two versions of guidance for MacOS users and Windows users, respectively. In our experience, a single version of guidance for all OSs is not adequate, minor differences among OSs matter in the introductory-level course. 
    
    \item We provide three ways to access guidance: PDF download, website, and video. The conventional way is to upload guidance to the course learning management system as PDF, but its format (PDF) has limited representations, therefore we add a website version of the guidance. Due to COVID-19, students might not be able to get enough in-person instruction, we also provide video guidance to improve student engagement. 
    
    \item We constantly collect students' feedback, and periodically improve and update the guidance.
    
\end{itemize}

\subsection{Hierarchical Indexing}

We provide detailed descriptions of tutorials, students follow tutorials step-by-step, and set up a programming environment without effort. However, students' preferences regarding tutorials vary with their programming backgrounds. 
Tutorials with a flat structure cannot satisfy all students' requirements. 
Students who prefer video tutorials may not follow the video pace from the beginning. A common scenario is to start with a verbal demonstration and refer to a video when problems occur. It is difficult to locate the corresponding video timestamp from text if tutorials are flat structured. In addition, students with some programming knowledge require more concise messages, e.g. have installed Python and VS Code, but require guidance on Python environment in VS Code, which fail to provide if tutorials are flat structured. For students who are already familiar with the environment setup, it is hard for them to extract key messages from flat-structured tutorials.

We design hierarchical indexing on tutorials to fulfill diverse requirements. The hierarchical indexing consists of three levels: 1) High-level: we provide an abstract with all key messages at the beginning of each tutorial, such as software names, download links, software versions, etc. 2) Medium-level: we segment each tutorial into several parts and summarize with subtitles. We list subtitles and link the corresponding positions in tutorials. Students can locate the information they need without effort. 3) Low-level: we build mappings between video and verbal demonstration, i.e., verbal demonstrations for each step within a subtitle are linking the corresponding video timestamps. Thus, students switch between verbal and video demonstrations when problems occur.

\section{Evaluation and Student Responses}


In this section we evaluate two objectives: 

\begin{itemize}
\item Validate VS Code is a suitable IDE for the introductory-level Python programming courses

\item Verify the value and necessity of VS Code's Guidance

\end{itemize}

\subsection{Course Description}
We introduce VS Code and VS Code guidance to a CS1 Python programming course in Fall 2022, the course enrollment is 141. This CS1 course is restricted for non-CS students, with an emphasis on basic programming skills and engineering applications. Students learn how to solve problems through writing Python programs, particular elements include: the development of Python programs from specifications; documentation and coding style; use of data types, control structures and data structures; abstractions and verification. 

The course uses a flipped classroom format. Every week, students start conceptual learning of the topics covered in the week by watching videos and completing self-check quizzes online. Then they attend a 100-minute class exercises session and a 165-minute lab exercises session to participate in the active learning activities. Students work on weekly homework assignments and project tasks after class to reinforce the knowledge and skills of the week. Students are required to run the in-class coding exercises via the Jupyter Notebook extension installed in VS Code, and complete the programming tasks (homework assignments, lab exercises, and projects) in VS Code by themselves. 




\subsection{VS Code and VS Code Guidance Evaluation}

We designed surveys with 21 questions, to collect students' feedback on the VS Code and VS Code Guidance at the end of the semester. The survey first collects students' backgrounds, then invites students to rate both VS Code and VS Code guidance based on their programming experience throughout the semester. In total, 42 valid responses were collected.

First, we collect students' background information. As a typical CS1 course, 79\% of students are freshmen and sophomores, 86\% of students did not take any programming course in the past, and 74\% of them have less than one year of coding experience.

Then students are invited to rate the VS Code and VS Code guidance performance throughout the semester. They rate VS Code from the following four aspects, the degree of satisfaction lies between 1 to 5, in which 1 is strongly dissatisfied and 5 is strongly satisfied.

\begin{itemize}
    \item Visual appeal: rate the VS Code's user interface, and the editor layout.
    \item Extension ecosystem: rate the way to search/install/uninstall extensions.
    \item debugging experience: rate the VS Code's built-in debugger, whether it helps accelerate students' edit, compile, and debug loop, and if the recommended debugger extensions are helpful. 
    \item editing experience: rate the overall editing experience of VS Code. If the basic editing features (i.e., keyboard shortcuts and Command Palette) are useful and beginner-friendly. Also rate IntelliSense, if code editing features such as code completion, parameter info, and content assist are helpful.
    
\end{itemize}

\begin{table}[H]
    \centering
    \caption{The averaged answers of 42 students' satisfaction over four aspects: visual appeal, extension ecosystem, debugging experience, and editing experience. The degree of satisfaction lies between 1 to 5, in which 1 is strongly dissatisfied and 5 is strongly satisfied.}
    \resizebox{\linewidth}{!}{
    \begin{tabular}{ c|c|c|c|c }
    \toprule
    Aspects & Visual Appeal & Extension & Debugging & Editing\\
    \midrule
    Average Answers & 4.17 & 3.81 & 4.02 & 4.05 \\
    \bottomrule
    \end{tabular}
    }
    \label{tab:VS1}
\end{table}

Based on the responses, 61\% of students do not have IDE or coding platform experience in the past, and 39\% of students have used some other IDEs such as Atom, Spyder, PyCharm/CLion/IntelliJ, etc. , As shown in Table~\ref{tab:VS1}, the average rating for visual appeal, extension ecosystem, debugging experience, and editing experience are 4.17, 3.81, 4.02, and 4.05, respectively.

The students' overall satisfaction with VS Code guidance is 4.2. With the help of detailed tutorials in VS Code guidance, 74\% students consider VS Code quite easy to install and use, and 76\% students consider VS Code as a good IDE to code with. There are 13 students who claim they had issues/trouble with VS Code guidance throughout the semester, and we have collected these issues and fixed/updated them in the VS Code guidance.

\subsection{Class Averages}

We also compare the class averages of Fall 2022 with Spring 2022, the results are shown in Table~\ref{tab:GPA}. In Fall 2022, the class average total score is 85.10\%, and the class average of all the coding assignments is 89.31\%, whereas 82.86\% and 88.37\% in Spring 2022, respectively.

\vspace{-0em}
\begin{table}[H]
    \centering
    \caption{The class average of coding assignments and total score in the Spring and Fall 2022 semesters.}
    \resizebox{0.7\linewidth}{!}{
    \begin{tabular}{ c|c|c }
    \toprule
    Class Average & Coding Assignment & Total Score\\
    \midrule
    Spring 2022 & 88.37\% & 82.86\%  \\
    \bottomrule
    Fall 2022 & 89.31\% & 85.10\%  \\
    \bottomrule
    \end{tabular}
    }
    \label{tab:GPA}
\end{table}

\vspace{-1em}
Although the comparison in Table~\ref{tab:GPA} shows that the class averages of Fall 2022 are higher than Spring 2022, we cannot simply accredit the average raise to VS Code or VS Code guidance. In Spring 2022, we used the traditional classroom, and students are required to use Spyder integrated within Anaconda, whereas we use the flipped classroom and VS Code in Fall 2022. But according to the student's experience, we can conclude that both VS Code and VS Code guidance have positive effects and are useful in completing coding assignments.

\subsection{Student Comments}
At the end of the student survey, students are invited to comment on their VS Code and VS Code guidance experience throughout the semester. One student described his/her overall experience as:
\begin{quote}
    I can view Python files and text files alike without even making a new window using VS Code. The interface is visually appealing. The VS Code tutorial is straightforward. Instructions are easy to follow. That's all a tutorial really needs, in my opinion.
\end{quote}

Another student especially liked the user interface of VS Code:
\begin{quote}
     I like how user-friendly VS Code is, the overall interface is easy to understand, allowing us to focus more on learning and applying Python concepts. 

\end{quote}

Some students favor the dark mode in VS Code, one student claimed it improves his/her coding experience:
\begin{quote}
    I like VS Code uses color coordination to make different code types and functions stand out. I also like that by default the background is black it makes it easier to read. 90\% of the time it's very easy to find mistakes within codes. Overall the tutorial is easy to follow. 
\end{quote}
We observe that many students prefer dark mode compared with bright mode, it may be because dark mode or dark theme stands out the syntax highlighting. With a light background, the code texts are mostly darker colors, while more colorful with a dark background. In addition, compared with pedagogical IDEs, changing themes in VS Code is quite easy. Students highly raise the theme extensions in VS Code, one student claimed:


\begin{quote}
    I like the way how VS Code installs extensions, I especially love theme extensions in Marketplace, whenever I am bored with codes I change my theme, and it just becomes a totally new software!
\end{quote}

Some students enjoy the VS Code guidance by simply saying: 
\begin{quote}
    I was pleased with how clear the steps were, and that in most steps there were screenshots to go along with the instructions. 
\end{quote}

We also find students like the hierarchical indexing in the VS Code guidance, one student commented on it as:
\begin{quote}
    It was simple to follow and the links were embedded into the steps. Also, there are pictures and screenshots to help guide you in each step, it helps me to prepare this material before classes.
\end{quote}

From the feedback, we collect many constructive suggestions for both VS Code and VS Code guidance. Some students complain about the complex commands on the terminal: 
\begin{quote}
    The terminal's instructions are complicated and hard to understand.
\end{quote}
It's true that we may involve plenty of terminal instructions at the beginning of class, but the instructor thinks those instructions/commands are required for this class. To fix this problem, we divide the command study into different labs and add detailed explanations on every instruction.  

Another student considered VS Code guidance useful, but suggested we could involve more helper extensions in class:

\begin{quote}
    Tutorials are thorough, pretty straight to the point. But in my opinion, VS Code lacks collaborative elements, especially in the lab. An extension that takes about 1 minute to install allows you and your lab partner to collaborate on the same Python file. My lab partner and I did this when we had to do debugging and coding exercises during the lab portion of the course. I feel that collaboration via a `shared' document would be beneficial because, without it, one user mostly does the code by themselves.
\end{quote}

The live share extension enables students to share screens and collaborate with classmates/TAs/instructors on the same code without the need to sync code or configure the same development tools, settings, or environment~\cite{liveshare}. We are evaluating the feature and will introduce it to students after careful consideration.

To summarize, students' overall experience with VS Code and VS Code guidance is positive and encouraging, which shows that VS Code and guidance are valuable and promising for the CS1 course.


\section{Discussions}

\myparabb{Discussions on education-related concerns.}
VS Code has some advanced features that raise some debates for their use in introductory-level programming courses, such as code auto-completion and function signature hints. Some argue that supplementary features would develop a blind dependence on IDEs rather than truly understanding the code. In practice, our teaching plan covers all necessary programming knowledge and concepts, and features like auto-completion and function signature hints help students learn faster and more efficiently. Moreover, these features can be customized or completely disabled in VS Code if required.

\myparabb{Discussions on some limitations.}
We can see some limitations in this study. We only evaluate an introductory Python course. 
The reason is that our department offers Python as the introductory programming language. However, VS Code supports multiple programming languages, such as Matlab, Java, and C/C++. It is straightforward to adapt our guidance to courses with other programming languages. We will partner with instructors in our department to report more VS Code experiences.

\section{Conclusions and Future Work}
This paper describes the experiences of introducing Visual Studio Code in an introductory (CS1) Python programming course at a big engineering university. In this paper, we investigate VS Code as a satisfactory IDE for CS1 programming courses. To better help students, we develop comprehensive VS Code guidance for students with various programming backgrounds. We perform evaluations among students and validate the practicality of VS Code and verify the quality of our VS Code guidance. We periodically update and improve the guidance with the collected feedback. We are now practicing VS Code and our VS Code guidance in more CS1 programming courses with programming languages besides Python.

Our future work is two-fold. First, we will continue exploring useful VS Code extensions available in the marketplace, which can benefit education. Second, we will develop useful VS Code extensions to further support education-related activities. Specifically, we will develop extensions that can better engage students when they are doing programming assignments.



\newpage

\bibliographystyle{ACM-Reference-Format}
\bibliography{iticse2023}


\end{document}